\documentstyle[12pt,epsfig]{article}
\newlength{\dinwidth}
\newlength{\dinmargin}
\newcommand{\ba}{\begin{array}}
\newcommand{\ea}{\end{array}}
\newcommand{\be}{\begin{equation}}
\newcommand{\ee}{\end{equation}}
\newcommand{\bea}{\begin{eqnarray}}
\newcommand{\eea}{\end{eqnarray}}
\newcommand{\gsim}{\mathrel{\mathop{\kern 0pt \rlap
  {\raise.2ex\hbox{$>$}}} \lower.9ex\hbox{\kern-.190em $\sim$}}}

\def\ben{\begin{equation}}
\def\een{\end{equation}}
\def\bea{\begin{eqnarray}}
\def\eea{\end{eqnarray}}
\def\nn{\nonumber}

\def\pg{2\pi g_s}

{\rm

\def\tbeta{{\tilde{\beta}}}
\def\rt{\gamma}

\begin{document}

\thispagestyle{empty}
\addtocounter{page}{-1}
\vskip-0.35cm
\begin{flushright}
UK/02-04 \\
{\tt hep-th/yymmxxx}
\end{flushright}
\vspace*{0.2cm}
\centerline{\Large \bf D branes in 2d string theory 
and Classical Limits}\footnote{Based
on talks at {\em ``Third International Symposium on 
Quantum Theory and Symmetries''}, Cincinnati, September 2003 and
{\em ``Workshop of Branes and Generalized Dynamics''}, Argonne,
October 2003.}

\vspace*{1.0cm} \centerline{\bf Sumit R. Das}
\vspace*{0.7cm}
\centerline{\it Department of Physics and Astronomy,}
\vspace*{0.2cm}
\centerline{\it University of Kentucky, Lexington, KY 40506 \rm USA}
\vspace{0.7cm}
\centerline{\tt das@pa.uky.edu}

\vspace*{0.8cm}
\centerline{\bf Abstract}
\vspace*{0.3cm}

In the matrix model formulation of two dimensional noncritical string
theory, a D0 brane is identified with a single eigenvalue excitation.
In terms of open string quantities (i.e fermionic eigenvalues) the
classical limit of a macroscopically large number of D0 branes has a
smooth classical limit : they are described by a filled region of
phase space whose size is $O(1)$ and disconnected from the Fermi sea.
We show that while this has a proper description in terms of a {\em single}
bosonic field at the quantum level, the classical limit is rather
nontrivial.  The quantum dispersions of bosonic quantities {\em
survive in the classical limit} and appear as additional 
fields in a semiclassical description.  This reinforces the
fact that while the open string field theory description of these
D-branes (i.e. in terms of fermions) has a smooth classical limit, a
closed string field theory description (in terms of a single boson)
does not.

\vspace*{0.5cm}

\newpage

\section{Overview}

D branes of string theory are usually described in terms of
excitations of open strings. In recent years it has become clear that
they also appear as classical solutions of open string field
theory. Their description in a closed string theory is, however, rather
obscure. Such a closed string description is desirable from
various points of view. In particular, we would like to understand
better how D-branes appear as states in a quantum theory of
gravity. We know that collections of large number of 
D-branes appear as classical solutions in
supergravity, most often with sources. In a complete theory there
cannot be any external source : in supergravity these sources of
course represent the effects of stringy degrees of freedom and we
should be able to do without them if we have a satisfactory off shell
description of closed strings.

In almost all cases of interest, however, we do not have a tractable
off shell description of closed strings. The most interesting
exception is noncritical string theory in two space-time
dimensions \cite{twodim}. 
Here, there is a complete nonperturbative description in
terms of the double scaling limit of matrix quantum mechanics - a
subject which has been vigorously pursued in the 1990's. It has
been long suspected - ever since the original work of 't Hooft - 
that theories of $N \times N$ matrices (e.g. gauge
theories) become, at large N, string theories. The new element which
was learnt in the 1990's is that these noncritical string theories typically
live in one more dimension \cite{noncritical,pola}. 
For matrix quantum mechanics the space of
eigenvalues of the matrix in fact provides the space dimension in
which the string moves \cite{dasjev}. 
This allows a non-gravitational theory in $0+1$ dimensions to be
dual to a theory which contains gravity in $1+1$ dimensions, thus
providing the earliest example of holographic correspondence. In this
case, the string theory is rather trivial : for the bosonic string the
only dynamical degree of freedom is a single massless scalar, which is
related to the collective field \cite{collective} - the density of
eigenvalues. Recall that the eigenvalues of this theory behave as
fermions and the collective field may be regarded as a bosonization of
these fermions \cite{fermions}, 
albeit the bosonization is rather complicated and
subtle. Thus we have a well defined closed string field theory as well
as a well defined holographic dual.

There were several disappointing features of the matrix model
description of two dimensional string theory. Even though gravity is
not dynamical in two dimensions, there can be nontrivial gravitational
backgrounds - in fact there is a black hole solution \cite{bhole}.
Despite considerable effort, no convincing description of black holes
emerged in the matrix model. Secondly, as is well known, the double
scaling limit becomes a theory of fermions moving in an inverted
harmonic oscillator potential. It was clear that the ground state
obtained by filling all the levels below the Fermi surface on one side
of the potential corresponds to the ground state of the bosonic
string, while small excitations represented a single massless
scalar. However this is a nonperturbatively unstable situation. The
stable situation is one obtained by filling both sides of the
potential : however the corresponding string theory for this was never
clear \cite{twosides}.  Thirdly after the discovery of D-branes in
usual critical string theory, it became clear that these are generic
states in {\em all} string theories. Indeed worldsheet descriptions of
D0 and D1 branes in the model were discovered \cite{liouvillebrane}.
However the matrix model description of these were not clear.

Recent progress has improved the situation considerably. In a very
interesting development, matrix quantum mechanics was reinterpreted as
a theory of $N$ D0 branes \cite{verlinone}. The worldline gauge field
forces all physical states to be singlets. A single unstable D0
brane is interpreted as a single eigenvalue excitation over the filled
Fermi sea, whose worldsheet description is a product of the D0
boundary state with Sen's rolling tachyon \cite{kms,verlintwo}.
The
model with only one side of the potential filled is the usual bosonic
string theory, while the model with both sides filled is the 0B theory
\cite{zerob}. In the latter case, the small excitations of the Fermi
level on both sides are linear combinations of the two massless
scalars in the perturbative spectrum. There is a related matrix model
for the Type 0A theory as well \cite{zeroa}.  Black holes, however,
continue to be a mystery. 
Single eigenvalue excitations have played an important role in the
subject, e.g. in the discovery that nonperturbative effects in string
theory behave as $e^{-1/g_s}$ rather than $e^{-1/g_s^2}$
\cite{shenker}. Single particle 
solitonic excitations also appear as states in exact
collective field theory as a separate branch \cite{jevickiexact}.
Explicit single particle states, particularly corresponding to tunnelling,
have been studied some time ago in
\cite{mende} and \cite{dmwclass}.

In the modern interpretation, the matrix degrees of freedom, or
equivalently fermions, are the open string degrees of freedom and the
collective field is the closed string degree of freedom. Open-closed
duality is then bosonization \cite{sen}. The open string description
is simple since the fermions are non-interacting. Near the hump of the
potential, the string coupling is large and the bosonic description is
complicated. However in the asymptotic region the bosons are free
massless particles. The late time decay product of a D0 brane can be
seen to agree with a coherent state of these massless particles, as
expected from the worldhsheet picture \cite{kms,kraus}.

This note deals with the nature of classical limit of D-brane states
from the point of view of closed string theory, largely based on work
done in collaboration with S.D. Mathur and P. Mukhopadhyay.

In the limit $g_s \rightarrow 0$ each of the fermions may be thought
of as particles moving classically in the inverted harmonic oscillator
potential, subject to the Pauli principle \cite{polchinski}. In the
phase space of fermions, the ground state is the filled Fermi sea. In
the bosonized theory, this limit corresponds to the classical limit of
the field theory. Thus, small deformations of the Fermi sea correspond
to classical waves of the bosonic theory with small amplitudes. The
classical dynamics of such small ripples is given by classical
solutions of collective field theory, which may be obtained directly
from the dynamics of the phase space density of fermions. The time
evolution of a single unstable D-brane is then represented by the
trajectory of a single fermion which is {\em disconnected} from the
Fermi sea. A state with many D0 branes is represented as a
disconnected blob of fermi fluid.  As we will soon see, and as has
been noted earlier \cite{dmwclass}, the resulting density of fermions
{\em do not} obey the classical equations of collective field
theory. In fact, the density of fermions $u(x,p,t)$ in phase space is
equivalent to an infinite number of fields in configuration space,
e.g. the various moments $\int dp p^n u(x,p,t)$ and for a generic
configuration these fields are independent. For quadratic profiles of
the Fermi surface, i.e. for profiles in which a constant $x$ line
intersects the Fermi surface precisely twice, these various moments
are not independent of each other and the configuration can be
determined by the collective field and its conjugate momentum
\cite{dmwbose}.  This would seem to indicate that the ``closed
string'' description of D0 branes necessarily involve an infinite
number of bosonic fields at the classical level.

We will argue, however, that at the full {\em quantum} level these
states are still described by a {\em single scalar field}, using a key
result of \cite{dmathur} about the physics of formation of folds on
the fermi surface.  If we start out with some ripple on the Fermi
surface - which is represented by a classical configuration of the
collective field - time evolution generically results in folds which
make the Fermi profile non-quadratic. In \cite{dmathur} it was argued
that this is still described as a state in a quantum theory involving
a {\em single} scalar field. What happens is that an initial coherent
state representing a classical wave evolves into a state in which
quantum dispersions become $O(1)$ rather than $O(\hbar)$ after a
certain time. This is precisely the time at which a fold forms on the
Fermi surface.

For D0 branes, the fermi surface has disconnected pieces to begin with
and the same phenomenon occurs. For a single D0 brane we find that
quantum flutuations of closed string quantities diverge in the
classical limit. This is entirely expected. Just as in critical string
theory, we do not expect that a state of a small number of
D-branes can be described as a classical configuration of closed string
fields.  On the other hand, one would expect that the state of a {\em
large} number of D0 branes (i.e. $O(1/g_s)$ in number) should be a
classical state of the closed string theory, i.e. the bosonic field
should have vanishing quantum dispersion in the $g_s \rightarrow 0$
limit.  We will show, however, that this is not true.  Quantum
fluctuations of quantities in the bosonic field theory do not vanish
in the $g_s \rightarrow 0$ limit and are in fact of the same order as
the expectation values of the fields themselves. Consequently, these
fluctuations enter the effective classical equations as independent
fields.
It
would be interesting to find the relationship between the description
of D0 brane states in terms of bosons as discussed above and the
solitonic states found in an exact treatment of collective field
theory in \cite{jevickiexact}.

The upshot of this is that while the open string field theory
description of D0 branes in this model has a smooth classical limit in
which Ehrenfest's theorem holds (as argued in \cite{sen}), a
description in terms of closed strings necessarily involves states
which have large quantum fluctuations.

\section{Preliminaries}

Matrix quantum mechanics is described by a Hamiltonian 
\ben
H = -{1\over 2\beta N}\sum_{ij} {\partial^2 \over \partial M_{ij}^2}
+ \beta N {\rm tr} V(M)
\label{eq:one}
\een 
where $V(M)$ is some potential for a $N \times N$ hermitian matrix $M$
which has a maximum at some value of $M$. We will work in the singlet
sector of the theory. In this sector one has the dynamics of $N$
eigenvalues which behave as fermions - so that one has a set of $N$
fermions in an external potential.
The double scaling limit is obtained by tuning the coupling $\beta$ to
some critical value such that the fermi level $\epsilon_F$ reaches the
maximum of the potential, $\epsilon_c$ and taking $N \rightarrow
\infty$ keeping
\ben
\mu = \beta N (\epsilon_c - \epsilon_F)
\label{eq:two}
\een
fixed. As is well known, in this limit only the quadratic part of the
potential survives. After a suitable rescaling of the space coordinate
$x$, the theory may be written in terms of a second quantized
fermionic field $\psi (x,t)$ with a Hamiltonian
\ben
H = \int dx [{1\over 2\mu} \partial_x \psi^\dagger \partial_x \psi -
{1\over 2} \psi^\dagger \psi]
\label{eq:three}
\een
This form of the hamiltonian clearly shows that the coupling constant
\ben
g_s = {1\over \mu}
\label{eq:four}
\een
acts like a $\hbar$ so that the classical limit is $g_s \rightarrow
0$.
The singlet sector states are best described in terms of the density
of fermions or the collective field $\rho (x,t)$. The effective theory
for $\rho$ can be obtained directly from the matrix hamiltonian
(\ref{eq:one}) by a change of variables to 
\ben
 \rho (x,t) = {\rm Tr} ~ \delta (x\cdot I - M(t)) 
\label{eq:five}
\een
with the result
\ben
H = {1\over 2}\int dx[{1\over 3}(P_+^3 - P_-^3) - (x^2 - 2\mu)(P_+-P_-)]
\label{eq:six}
\een
where $P_\pm$ are
\ben
P_\pm = \partial_x \Pi_\rho \pm \pi \rho
\label{eq:seven}
\een
Here $\Pi_\rho$ is 
the momentum conjugate to $\rho$. The classical solution is given by 
\ben
\Pi = 0 ~~~~~~~~\rho = {1\over \pi}P_0 = {1\over \pi}{\sqrt{x^2 - 2\mu}}
\label{eq:eight}
\een
Expanding around the classical solution as
\ben
P_\pm = \pm P_0 + {1\over {\sqrt{2}}~P_0}[\Pi_\phi \pm \partial_q \phi]
\label{eq:nine}
\een
where we have defined a new space variable $q$ by
\ben
\partial_q = P_0 \partial_x
\label{eq:ten}
\een
one can then easily see that the quadratic part of the hamiltonian is
\ben
H_2 = {1\over 2}\int dq [\Pi_\phi^2 + (\partial_q \phi)^2]
\label{eq:eleven}
\een
which identifies $\phi$ as a massless scalar field and $\Pi_\phi$ its
canonically conjugate momentum.

On the left part of the potential one has
\ben
x = -{\sqrt{2\mu}}\cosh q
\label{eq:thirteen}
\een
However this choice of a space coordinate is valid only in the
classically allowed region. For the region inside the potential hump
one has to use a different parametrization. A description of
nonperturbative physics is of course best done in terms of the
fermionic theory formulated in the full $x$ space.

\section{Classical bosonization}

At the classical level, the collective Hamiltonian can be alternatively
obtained by considering the phase space dynamics of the fermions. Let
$u(x,p,t)$ denote the density of fermions in phase space. In the
classical limit $g_s \rightarrow 0$, Pauli principle dictates that $u =
{1\over 2\pi g_s}$ in regions where the fermions are present and $u=0$
elsewhere. The ground state is thus given by
\ben
 u(p,x,t) = {1\over \pg} \theta (x^2-p^2-2\mu)
\label{eq:fourteen}
\een
which corresponds to the Fermi sea filled on both sides of the
potential. Consider now a small fluctuation which may be thought of a
small ripple on the Fermi sea. We will first restrict our attention to
perturbations which produce a {\em quadratic} profile. This means that
a $x$= constant line intersects the Fermi surface $x^2-p^2 = 2\mu$
precisely {\em twice}, at $P_\pm (x,t)$. In this special case, the
density of fermions in space at some given time, $\rho (x,t)$, may be
obtained as \cite{polchinski}
\ben
\rho (x,t) = \int {dp}~u(x,p,t) = {1\over \pg} 
[P_+(x,t)-P_-(x,t)]
\label{eq:fifteen}
\een
while the momentum density $\nu (x,t)$ is given by
\ben
\nu (x,t) = \int {dp}~p~u(x,p,t) = {1\over \pg}{1\over
  2}[P_+^2-P_-^2]
\label{eq:sixteen}
\een
Identifying $\nu = \rho \partial_x \Pi$ one gets the standard
relationship between collective field
(and its conjugate momentum) with $P_\pm$. The hamiltonian
(\ref{eq:six}) then follows by noting that
\ben
H = \int {dx dp}{1\over 2}[p^2-x^2 + 2\mu] u(x,p,t)
\label{eq:seventeen}
\een

\begin{figure}[ht]
   \vspace{0.5cm}
\centerline{
   {\epsfxsize=8.5cm
   \epsfysize=8cm
   \epsffile{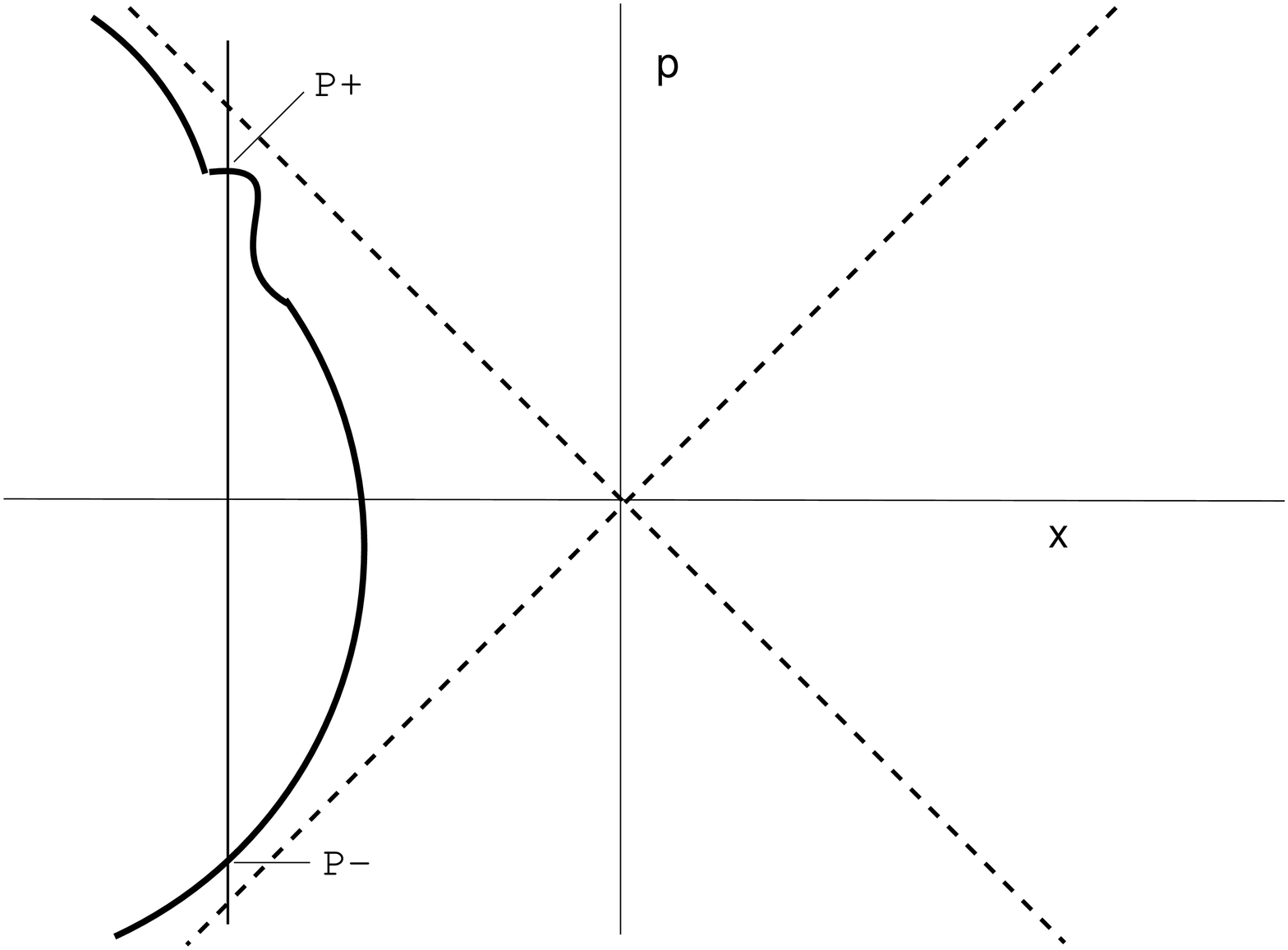}}
}
\caption{\sl A quadratic profile of the Fermi surface}
\label{quadratic}
\end{figure}

As is well known, however, the quadratic profile approximation made
above is valid only for very small and shallow ripples on the fermi
sea. A generic intial quadratic deformation would inevitably form
``folds'' rendering the profile non-quadratic. 

Of particular interest are initial states which correspond to
non-quadratic profiles. Single eigenvalue excitations or excitations
of a bunch of eigenvalues are examples of these since the filled
fermi sea now has disconnected pieces. The classical dynamics is still
described by the equation of motion for the phase space density
$u(x,p,t)$ 
\ben
(\partial_t + x\partial_p + p\partial_x) u(x,p,t) = 0
\label{eq:eighteen}
\een
and a constraint which implements Pauli principle.
However this equation cannot be any longer reduced to classical
collective field theory equations \cite{dmwclass}. A complete solution
of $u(x,p,t)$ corresponding to the rolling tachyon has been
obtained recently in \cite{mwclass}.

This seems to indicate that at the classical level there are an
infinite number of bosonic fields since we can define an independent
bosonic field by taking some moment of $u$. A useful definition of the
fields is provided by the following \cite{avanjevicki}
\bea
\int {dp}~u (x,p,t) & = & {1\over \pg}[\beta_+(x,t) -
  \beta_-(x,t)]
 \nn \\
\int {dp}~p~u(x,p,t) & = & {1\over \pg}[{1\over
    2}(\beta_+^2 (x,t) - \beta_-^2 (x,t)) + (w_{1+}(x,t) -
  w_{1-}(x,t))] \nn \\
\int {dp}~p^2~u(x,p,t) & = & {1\over \pg}[({1\over
    3}\beta_+^3 (x,t) + \beta_+w_{1+}(x,t) + w_{2+}(x,t)) \nn \\
& & ~~~~~~-({1\over
    3}\beta_-^3 (x,t) + \beta_-w_{1-}(x,t) + w_{2-}(x,t))]
\label{eq:nineteen}
\eea
The $\beta_{\pm},w_{n\pm}$ satsify $w_\infty$ algebras under usual
Poisson brackets. 

In this parametrization, the presence of nonzero
$w_n$ signifies that the profile is not quadratic. For example if the
fermi sea consists of two disconnected pieces  (Fig.2) and a given $x$ =
constant line intesects the fermi surface at four points $P_{i\pm}$
and $P_{2\pm}$ one has
\bea
\beta_+ & = & P_{1+}-P_{1-}+P_{2+} \nn \\
\beta_- & = & P_{2-} \nn \\
w_{1+} & = & (P_{1+}-P_{1-})(P_{1-}-P_{2+}) \nn \\
w_{1-} & = & 0
\cdots
\label{eq:twenty}
\eea
The equations of motion for these various fields $\beta_\pm (x,t), 
w_{\pm i}(x,t)$ (as usual half of them should be regarded as conjugate
momenta) may be obtained from (\ref{eq:eighteen}). These equations
couple the $w_{\pm i}$ with the $\beta_{\pm i}$. As a result, even if
we start at $t=0$ with all the $w_{\pm i}=0$ time evolution may result
in a nonzero $w_{\pm i}$ at a later time. This is the physics of fold
formation from unfolded configurations.

We will be interested in rolling tachyon states representing decaying
D0 branes.  The semiclassical description of a single $D0$ brane at
some instant of time is a filled region in phase space of size
$O(g_s)$ which is {\em disconnected} from the fermi sea, just like in
Fig 2. Equation (\ref{eq:twenty}) then implies that $w_{1+}$ is of
$O(g_s)$ and hence subleading in the classical limit. Note that each
of $P_{1\pm}$ are of $O(1)$ so that this state has an energy of order
$O(1/g_s)$. As expected a single $D0$ brane is a highly quantum state.

However, when we have a macroscopically large number of $D0$ branes we
expect a classical state in terms of bosons.  Indeed, in this case the
disconnected region has a size of $O(1)$, the various $w_i$ are also
of $O(1)$ and contribute to the classical limit.

In terms of an open string (fermionic) description this collection of
$D0$ branes has a smooth classical limit since the phase space density
$u(x,p,t)$ has small quantum fluctuation, implying that in phase space the
boundaries of the filled region are sharp.  In terms of a closed
string (bosonic) description we also have a classical limit. However
it appears that such a state necessarily involves an infinite number
of fields.

\begin{figure}[ht]
   \vspace{0.5cm}
\centerline{
   {\epsfxsize=8.5cm
   \epsfysize=8cm
   \epsffile{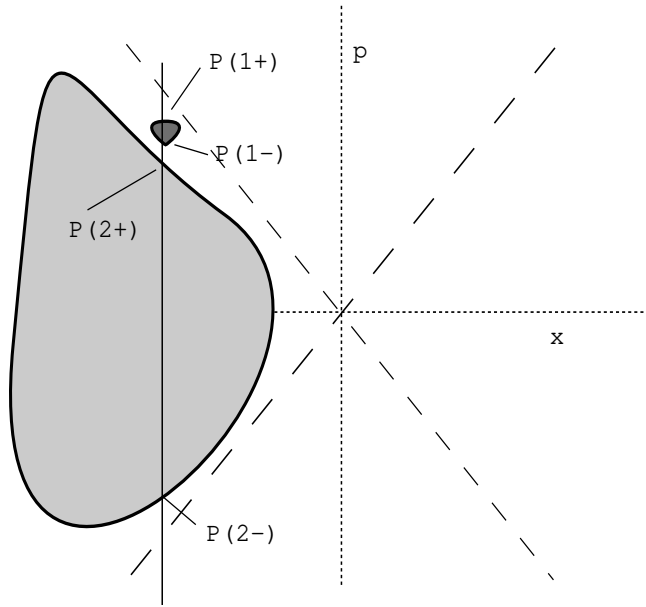}}
}
\caption{\sl A semiclassical description of D0 branes}
\label{dobrane}
\end{figure}

\section{Quantum bosonization}

\begin{figure}[ht]
   \vspace{0.5cm}
\centerline{
   {\epsfxsize=8.5cm
   \epsfysize=8cm
   \epsffile{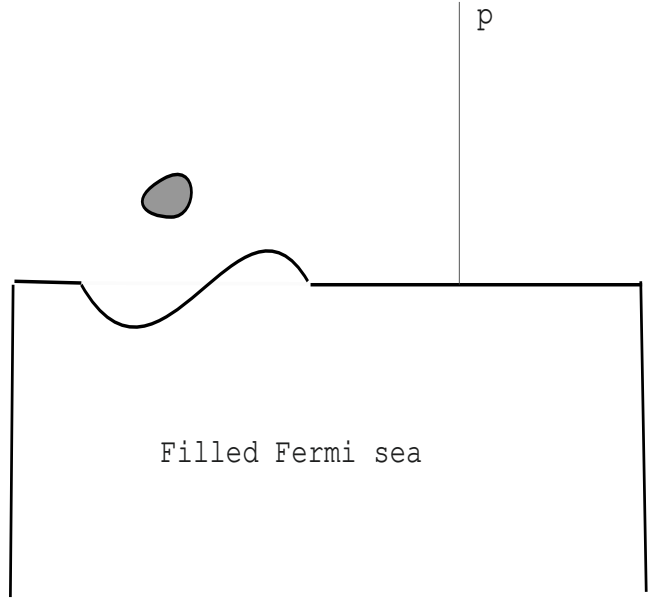}}
}
\caption{\sl An excitation of a set of right moving relativistic fermions}
\label{relativistic}
\end{figure}

The conclusion of the preceding section is puzzling. The discussion of
classical bosonization using phase space can be repeated for {\em
  relativistic} fermions as well. Consider for example the right
moving part of a massless relativistic fermion. Now the fermi sea has
only one edge - all states with negative momentum are filled. This in
fact simplifies the discussion of classical bosonization. Instead of
having two copies of each field labelled by $\pm$ we have a single
copy. For a state with some disconnected filled region, we have Figure
(\ref{relativistic}) instead of Figure (\ref{dobrane}).

The formulae are the same as in (\ref{eq:nineteen}) except that we
have to set $\beta_-=w_{i-}=0$ and rename $\beta_+ \rightarrow \beta$
and $w_{i+} \rightarrow w_i$. 
\bea
\int {dp}~u (x,p,t) & = & {1\over \pg}~\beta (x,t)
 \nn \\
\int {dp}~p~u(x,p,t) & = & {1\over \pg}[{1\over
    2}\beta^2 + w_1(x,t)] 
\label{eq:nineteena}
\eea
and so on.
The {\em dynamics} is of course
different, governed by a hamiltonian 
\ben
H_{rel} = \int {dp dx} ~p~u(x,p,t)
\een
This in fact ensures that the equations for $\beta$ and $w$ do not mix
with each other, 
so that if we have a state with $w_i = 0$ at some time, $w_i$'s
remain zero at all future times. Thus folds cannot form out of
unfolded configuration - a simple consequence of the fact that the
velocity of all fermions is the speed of light and independent of the
energy.

However, we can certainly have a state which corresponds to Figure
(\ref{relativistic}) to begin with, and we would conclude - as in the
previous section - that we need an infinite number of bosonic fields to
describe this classically. On the other hand we certainly have an
exact bosonization at the full quantum level in terms of single
(chiral) boson. Consequently all states
            of the theory can be written down exactly in terms of this {\em
  single} boson. The puzzle is : how is it that one needs an infinite
number of fields when one passes to the classical limit ?

\subsection{Bosonization at the algebraic level}

To see this let us review the bosonization formulae at an
algebraic level \cite{bosoni}. 
Suppose there are an infinite set of (Schrodinger
picture) fermionic operators $\psi_n $ and $\psi^\dagger_n $, with $n$
a positive or negative integer or zero, which satisfy the
anticommutation relations
\ben
\{ \psi^\dagger_m , \psi_n  \} =  \delta_{mn} ~~~~~~~~
\{ \psi^\dagger_m  , \psi_n  \}  = 0 ~~~~~~~~~
\{ \psi_m  , \psi_n  \} = 0
\label{eq:twotwo}
\een
Suppose we also have a vacuum state $|0>$ which obeys 
\ben
\psi_n |0>  =  0  ~~{\rm for}~n > 0 ~~~~~~~~~~~
\psi^\dagger_n |0> =  0~~{\rm for}~ n \leq 0
\label{eq:twothree}
\een
This allows the definition of a normal ordering prescription
\bea
: \psi^\dagger_m \psi_n : & = \psi^\dagger_m \psi_n~~~~& n > 0 \nn \\
: \psi^\dagger_m \psi_n : & = -\psi_n \psi^\dagger_m~~~~& n \leq 0
\label{eq:twofour}
\eea
Then the following bosonic operators
\ben
\alpha_m = \sum_n : \psi^\dagger_{-m+n} \psi_n :
\label{eq:twofive}
\een
satisfy the commutation relations
\ben
[ \alpha_m , \alpha_n ] = m \delta_{m+n,0}
\label{eq:twosix}
\een
Note that the above facts do not depend on the dynamics of the theory.

The simplest application of the above algebraic relation is
bosonization of a single chiral fermion into a single chiral boson. In
this case the integers $m,n$ may be regarded as labels for momenta in
a box. Then the operators $\psi_n$ may be combined into a single
chiral field $\psi (x)$
\ben
\psi (x)  = {1\over {\sqrt L}}\sum \psi_m~
e^{{2 \pi i m x \over L}} ~~~~~~~~
\psi^\dagger (x)  =  {1\over {\sqrt L}}\sum \psi^\dagger_m~
e^{-{2 \pi i m x \over L}}
\label{eq:twoseven}
\een
while the oscillators $\alpha_n$ can be combined
into a single chiral boson field $\phi (x)$
\ben
\partial_x \phi (x) = {1\over  L}
\sum \alpha_m ~e^{{2\pi i mx \over L}}
\label{eq:twoeight}
\een
The bosonization relationship then becomes
\ben
\alpha (x) \equiv \partial_x \phi (x) = : \psi^\dagger (x) \psi (x) :
\label{eq:twonine}
\een

The vacuum $|0>$ is the Fock vacuum of a free theory. We have chosen
the fermion to be right moving so that the energy of a single free
fermion is $E = p$, where $p$ is the momentum. The vacuum chosen above
then corresponds to the standard filled Dirac sea. In the $L
\rightarrow \infty$ limit the momentum becomes continuous. The above
relations have the standard limits. $\psi^\dagger (k)$ is a creation
operator for $k > 0$ and an annihilation operator for $k \leq 0$.

The fermion field may be also expressed in terms of the bosonic field
by the relationship
\ben
\psi (x) = : e^{i\alpha (x)} :
\label{eq:twoninea}
\een
where normal ordering in the bosonic theory is standard, i.e. all the
oscillators with negative indices have to be pushed to the right.
Using (\ref{eq:twoninea}) one can express other fermion bilinears in
terms of $\alpha$. One such relation which will be crucial in what
follows is
\ben
-{i\over 2}:[\psi^\dagger \partial_x \psi - (\partial_x
  \psi^\dagger)\psi]: = \pi : \alpha^2 (x):
\label{eq:aone}
\een

\subsection{Relativistic fermions and nonrelativistic fermions in
an inverted harmonic oscillator potential}

The key ingredient in the above is the presence of an {\em infinite}
filled Fermi (or Dirac) sea. This is the reason why the above
bosonization does not work for free non-relativistic fermions. The
energy of a single fermion is $h = {1\over 2m}p^2$ and now the
momentum space Fermi sea has a {\em upper} and a {\em lower} edge. For
example the fermionic modes $\psi (k)$ are creation operators for $
-k_F < k < k_F$ where $k_F$ is the Fermi momentum. It is now clear
that the momentum modes of the field $\alpha (x) = : \psi^\dagger \psi
:$ do not satisfy the standard bosonic oscillator algebra. It is only
for momenta {\em close to the Fermi sea} that we can use the standard
bosonization formulae. But this is simply because for such momenta
nonrelativistic fermions become effective relativistic with an
effective velocity of light $(k_F / m)$.

The situation is, however, quite different for nonrelativistic
fermions in an invereted harmonic oscillator potential. In this case
the fermion field may be expanded in terms of Parabolic cylinder
functions $\chi^\pm_\nu (x)$ with a continuous index $\nu$ which runs
from $-\infty$ to $+\infty$.
\ben
\psi (x) = \int d\nu [\psi^+_\nu \chi^+_\nu (x)
+ \psi^-_\nu \chi^-_\nu (x)]
\label{eq:thirty}
\een
where the modes $\psi_\nu$ satisfy the standard fermionic
anticommutation relations. The functions $\chi^\pm_\nu (x)$ are
eigenstates of the single particle Hamiltonian
\ben
h = {1\over 2} [p^2 - x^2]
\label{eq:threeone}
\een
The vacuum is defined by $\psi_\nu |0> = 0$ for $\nu > -\mu$ and
$\psi^\dagger_\nu |0> = 0$ for $-\infty <\nu < -\mu$. Thus the Fermi
sea is infinite, pretty much like a relativistic fermion. A
bosonization based on the exact eigenstates have been recently used in 
\cite{deboer} to investigate interactions involving D0 branes.

At the classical level this is clear from the fact that there is a
canonical transformation in phase space which maps the problem to that
of a relativistic fermion. For negative energy orbits on the left side
of the potential is this achieved by
\ben
x = - {\sqrt{2\xi}}\cosh~\eta ~~~~~~p = -{\sqrt{2\xi}}\sinh~\eta
\label{eq:threetwo}
\een
while for positive energy orbits we have a similar
paramertization. Here $-\infty < \eta < \infty$ and $0<\xi <
\infty$. In this parametrization the single particle Hamiltonian
becomes $h = -\xi$ and the Fermi level is $\xi = \mu$. Here $\eta$ is
a coordinate and $\xi$ is a momentum. In the ground state the filled
Fermi sea is $\mu < \xi < \infty$.

In any case, this shows that there is an {\em exact} bosonization
for fermions in an inverted harmonic oscillator potential in terms of
a {\em single} bosonic field since the filled Fermi sea is
infinite. The natural position space is of course not
$x$, but the conjugate of the quantum number $\nu$. 
In the semiclassical level this may be taken to be $\eta$. This is
clear from the fact that if we regard $x$ as the coordinate there is
an upper as well as a lower edge of the Fermi sea. When $\eta
\rightarrow \pm \infty$ motion in $\eta$ becomes the same as motion in
$x$ so that the bosonization yields a local field in $x$ space.  

Such a canonical transformation does not exist for {\em free}
non-relativistic fermions. or for non-relativistic fermions in a
generic potential where there is both a lower and upper edge of the
fermi sea.

\section{Disconnected fermi fluids and quantum fluctuations}

Let us return to the question posed in the beginning of the previous
section : how is it that at the classical level one finds that there
has to be an infinite number of bosonic fields ? Our discussion of
the previous section shows that for the present purpose we can treat
free relativistic fermions and non-relativistic fermions in a $-x^2$
potential in a unified way. Let us denote the coordinates in phase
space by $p$ and $q$.  For relativistic fermions the coordinate $q$ is the
standard position $x$ , while for the fermions coming from matrix
model $q$ stands for the conjugate to the quantum number $\nu$ - which
may be considered to the coordinate $\eta$ of (\ref{eq:threetwo}) at
the semiclassical level. With this understanding the bosonization
formulae are given by equations (\ref{eq:twonine} - \ref{eq:aone})
with the variable $x$ replaced by $q$. Our discussion below has
nothing to do with the dynamics of the theory. 

Consider now a Schrodinger picture 
state $|\Psi,t>$ of the system which is represented at the
semiclassical limit by a single disconnected
blob of fermi fluid on top of the infinite filled bulk of the fermi
sea, as depicted in Figure 3. Note that the boson $\alpha (q)$ is simply
the density of fermions in $q$ space, 
while the left hand side of (\ref{eq:aone})
is the {\em momentum density} of fermions. Therefore we can combine
the operator statements in (\ref{eq:twonine} - \ref{eq:aone}) with the
phase space definitions of semiclassical fields in (\ref{eq:nineteena})
to obtain at some given time
\ben
<\Psi,t|:\psi^\dagger (q)\psi (q) :| \Psi,t >
 =  <\Psi,t|\alpha (q) |\Psi,t> = 
\int {dp}~u (q,p,t)  =  {1\over \pg}~\tbeta (q,t) 
\label{eq:nineteend}
\een
\bea
-{ig_s \over 2}<\Psi,t| :[\psi^\dagger \partial_x \psi - (\partial_x
  \psi^\dagger)\psi]: |\Psi,t> & = & \pi g_s <\Psi,t |: \alpha^2(q) :
|\Psi,t> \nn \\
& = &  \int {dp}~p~u(x,p,t) =  {1\over \pg}[{1\over
    2}\tbeta^2 + w_1(q,t)] 
\label{eq:nineteenb}
\eea
where $\tbeta$ denotes the quantities where the ground state values
have been subtracted.

{\bf This immediately shows that $w_1$ is related to the quantum
  dispersion of the field $\alpha (q)$}
\ben
 <\Psi,t|: \alpha^2 (q) : |\Psi,t> - <\Psi,t|\alpha (q) |\Psi,t>^2
= {1\over 2\pi^2 g_s^2} w_1 (q)
\label{eq:threethree}
\een
A state with $w_1 \sim O(1)$ (as in Fig. 3) therefore has a rather
unconventional property ; the quantum dispersion $\Delta \alpha$ in
such a state is of the same order as the value of $<\alpha>$.

This state represents a large number of fermions which are all excited
above the fermi level leaving a gap. From the point of view of
fermions, such a state certainly has smooth semiclassical limit since
the position and momentum of each fermion is determined upto
$O(g_s)$. This means that the quantum spread of $u(p,q,t)$ is
controlled as usual by $g_s$ and vanishes in the $g_s \rightarrow 0$
limit. However quantities which are natural to the bosonized theory,
like expectation values of $\alpha$ have quantum dispersions which do
not vanish in the $g_s \rightarrow 0$ limit. Furthermore these quantum
dispersions like $<\alpha^n> - <\alpha>^n$ are precisely the extra 
infinite set of fields $w_n$ which appear in a semiclassical
description of the state in terms of bosons.

In contrast, a state with no diconnected pieces or folds - a smooth
shallow ripple - is a coherent state of the field $\alpha$ for which
dispersions of operators constructed from $\alpha$ vanish in the $g_s
\rightarrow 0$ limit. From the point of view of the bosonic theory
these are ``classical states''. However a
state with no folds generically time evloves into a state with folds
if the dynamics is appropriate. In \cite{dmathur} it was shown 
(for free nonrelativistic fermions)
that
this happens because the dispersions of $\alpha$, while initially of
$O(g_s)$ become $O(1)$ after a certain time. This time is precisely
the time when a fold forms.

It is now clear why states with non-quadratic profiles will not evolve
according to the classical equations of motion of the bosonic field
theory even in the semiclassical limit. The reason is that while there
could be an exact quantum bosonization, the states being considered
are not those which have a smooth classical limit for natural bosonic
variables. In other words Ehernfest's theorem does not hold for the
equations of motion of the closed string field theory, which is the
quantum bosonic theory in this case. Ehernfest's theorem of course
holds for the equations of motion of the open string field theory,
which is the theory of eigenvalues, simply because in the classical
limit the fermions have trajectories in phase space with a fuzziness
which is at most of order $g_s$.

\section{Explicit calculations}

In this section we will perform simple calculations which illustrate
the above point. 
The system we consider is the infinite set of
fermionic oscillators as in equations
(\ref{eq:twotwo})-(\ref{eq:twosix}). As argued in the previous section,
this system can refer to either a single chiral relativistic fermion
or nonrelativistic fermions which arise in the $c=1$ matrix model or
in fact any theory of fermions where the gorund state consists of an
infinite Fermi sea. The meaning of the integer index is of course
different in each case. With this understanding we can then use the
bosonization formulae (\ref{eq:twoseven} - \ref{eq:threeone}) where
$x$,  now denoted by $q$, is the position coordinate conjugate to the quantum
number $\nu$ which characterizes the eigenstates in the inverted
harmonic oscillator potential, where we have put this conjugate
position in a box of size $L$, thus rendering $\nu$ integral. In the
semiclassical limit this is the coordinate $\eta$ defined in
(\ref{eq:threetwo}).

In this section we consider properties of states at some given time $t$.
We calculate expectation values of various bosonic quantities in such
states, go over to the classical limit and compare them with the
expressions based on the phase space analysis presented above.

A ripple on the Fermi sea is a coherent state of the bosonic field
\ben
|C(q)> = {\cal N}~\prod_{n=1}^\infty 
e^{{C_n \alpha_{-n} \over 2\pi n g_s}} |0>
\label{eq:threeeight}
\een
where the $C_n$ are the fourier components of the function $C(q)$
\ben
C(q) = {1\over {\sqrt{L}}}\sum_n C_n ~e^{{2\pi i n q \over L}}
\label{eq:threenine}
\een
and ${\cal N}$ is a normalization factor which ensures that
$<C(q) | C(q)> = 1$.
Since $|C(q)>$ is an eigenstate of $\alpha_n$ for $n > 0$ with eigenvalue 
$C_n/2\pi g_s$ it is clear that
\ben
<C(q) | \alpha (q) |C(q) > = {1\over 2\pi g_s }C(q)~~~~~~~~~~ 
<C(q) | : \alpha^2 (q) : |C(q) > = {1\over 4\pi^2 g_s^2 }C^2(q)
\label{eq:forty}
\een

Comparing with (\ref{eq:nineteend})-(\ref{eq:nineteenb}) it is clear
that we have $\tbeta (q) = C(q)$ and $w_i(q) = 0$. This then
corresponds to a quadratic profile. In fact the state $|C(q)>$ has
been chosen so that the quantum dispersions of all bosonic operators
which do not involve $g_s$ explicitly vanish in the $g_s \rightarrow
0$ limit. These ``ripple'' states thus have smooth classical
limits from the point of view of the bosonic theory.

Consider now a state consisting of $M$ D0 branes, each of which are in
a definite energy eigenstate, and hence delocalized in $q$. The
D0 branes occupy successive energy levels between $n=n_A (> 0)$ and
$n=n_B=n_A+M-1$
\ben
|n_A,M> = \psi^\dagger_{n_A}\psi^\dagger_{n_A+1}\cdots \psi^\dagger_{n_B}~|0>
\label{eq:fourone}
\een
Let $\rt_n$ denote the fourier modes of $T(q) \equiv : \alpha^2 (q) :$
\ben
T(q) = {1\over L^2}\sum_n \rt_n ~ e^{-{2\pi i n q \over L}}
\label{eq:fourtwo}
\een
The following relationships then follow from the 
basic bosonization relation (\ref{eq:twofive})
\ben
[ \alpha_n , \psi^\dagger_m ]  =   \psi^\dagger_{-n+m}
~~~~~~~~ [\alpha_n, \psi_m ] = - \psi_{n+m}
\een
\ben
[ \rt_n , \psi^\dagger_m ]  =  (2m-n) \psi^\dagger_{-n+m}~~~~~~~~ 
[ \rt_n, \psi_m ] = - (2m+n) \psi_{n+m}
\label{eq:fourthree}
\een
Note that the ground state obeys
\ben 
\alpha_n |0> = 0~~~~~~ \rt_n |0> = 0~~~~~~~{\rm for~} n \geq 0
\label{eq:fourfour}
\een
These relations lead to the following results
\bea
<n_A,M|\alpha (q)|n_A,M> & = & {M \over L} \nn \\
<n_A,M|: \alpha^2 (q) : |n_A,M> & = & {M \over L^2}(2n_A + M-1)
\label{eq:fourfive}
\eea
The result for $<\alpha>$ simply reflects the fact that we have $M$ fermions
which are completely delocalized in $q$ space. 

The physical momenta in $q$ space corresponding to the lowest and the
highest filled level above the Fermi sea are given by
\ben
p_A = {2\pi g_s n_A \over L}~~~~~~~~p_B = {2\pi g_s (n_A + M - 1) \over L}
\label{eq:foursix}
\een
In terms of these quantities
\bea
<n_A,M|\alpha (q)|n_A,M> & = & {1\over \pg} (p_B-p_A) + 
{1\over L} \nn \\
<n_A,M|: \alpha^2 (q) : |n_A,M> & = & {1\over 4\pi^2 g_s^2}(p_B^2 - p_A^2)
+ {1\over \pg L}(p_A + p_B)
\label{eq:fourseven}
\eea

In the limit $L \rightarrow \infty$ the physical momenta become
continuous. Furthermore in the classical limit $g_s \rightarrow 0$ the
momenta become $O(1)$ if $n_A \rightarrow \infty$ keeping $p_A, p_B$
fixed. In this limit we can ignore the $O(1/L)$ terms and the above
results reproduce the expectations from classical Fermi fluid
picture. In particular, the quantities defined in
(\ref{eq:nineteend}-\ref{eq:nineteenb}) we have
\bea
\tbeta & = & (p_B - p_A) \nn \\
w_1 & = & (p_B - p_A) p_A
\label{eq:foureight}
\eea
as expected.

When the number of D0 branes $M \sim O(1)$, we have $w_1 \sim {g_s
\over L}$ so that the quantum dispersion of $\alpha$ is large
\ben
{\Delta \alpha \over \alpha} \sim {{\sqrt{w_1}}\over \tbeta} \sim
{\sqrt{{Lp_A \over g_s M}}}
\label{eq:fournine}
\een
This clearly diverges in the classical limit $g_s \rightarrow
\infty$. This is entirely expected. As in critical string theory a
state of a finite number of D branes is a highly quantum state from
the point of view of closed string theory - we have simply
reproduced this result.

In critical string theory, however, closed string theory provides a
classical description of a large number of D-branes in the limit $g_s
\rightarrow 0$ and $M \rightarrow \infty$ with $g_s M$ held
fixed. This is the limit in which the collection of D branes is
described in terms of a classical gravitational background. In the
present case this corresponds to the limit in which the quantity $(p_B
- p_A) \sim O(1)$. In the $L \rightarrow \infty$ limit this is
described by a continuous disconnected region in the Fermi fluid
picture. Since we have considered D0 branes in energy eigenstates,
this is a band rather than a blob. Contrary to what one might have
expected, this corresponds to a limit in which the quantum dispersions
of the bosonic field ${\Delta \alpha \over \alpha} \sim O(1)$ rather
than $O(g_s)$ ! A classical description of such configurations of a
large number of D0 branes involves an infinite number of fields, while
an exact quantum description involves a single bosonic field.  The
additional fields are nothing but quantum dispersions which survive
the classical limit. In other words, while an open string description
of D0 branes has a smooth classical limit in accordance with
Ehernfest's theorem \cite{sen}, a closed string description does
not.

In order to describe disconnected blobs in phase space corresponding
to a large number of D0 branes approximately localized in space at any
given time we need to consider wavepackets of individual fermions. 
For example consider the single fermion states (in the $L \rightarrow
\infty$ limit)
\ben
|p_0,q_0> = \int {dk \over 2\pi}~{\rm exp}~[-{1\over 2}g_s (k-{p_0
    \over g_s})^2
  +ikq_0]~ |k>
\label{eq:wone}
\een
where 
\ben
|k> = {1\over {\sqrt{L}}} \psi^\dagger (k) |0>
\een
and the oscillators $\psi (k), \psi^\dagger (k)$ are related to the
oscillators with corresponding discrete index $n = {L \over 2\pi}k$ by
the relation $\psi (k) = {\sqrt L} \psi_n$. 

For $p_0 >> 0$ (corresponding to the average energy far above the Fermi sea),
and small $g_s$ 
the above results may be used to see that \footnote{The integral in
  (\ref{eq:wone}) is effectively from $k=0$ to $k=\infty$. However for
  $p_0$ large and $g_s$ small we may replace this by an iuntegral over
  the entire range}  
\ben
< \alpha (q) >  = { <p_0,q_0| \alpha (q) |p_0,q_0> \over
  <p_0,q_0|p_0,q_0>} = {1\over {\sqrt{\pi g_s}}}~e^{-{(q-q_0)^2 \over
    g_s}} 
\label{eq:wtwo}
\een
while
\ben
< : \alpha^2 (q) : > = {p_0 \over {\sqrt{\pi g_s}}}~e^{-{(q-q_0)^2 \over
    g_s}} 
\label{eq:wthree}
\een
in agreement with the results from the phase space picture.

To describe a blob of finite size, corresponding to a large number of
D0 branes localized approximately in phase space one can proceed with
states which are direct products of states described above.
It
is clear from the preceeding discussion that the conclusion remains
unchanged for such configurations, though the calculations are more
complicated. These calculations are useful in an understanding of
gravitational effects of D-branes \cite{progress}.

\section{Acknowledgements}

I would like to thank the organizers of QTS3 and
{\em ``Workshop on Branes and Generalized Dynamics''} 
for their invitation to present this talk.  
I also thank Samir D. Mathur and Partha
Mukhopadhyay for discussions and collaboration, and A. Jevicki,
I. Klebanov and S.R. Wadia for discussions. This work was supported by
National Science Foundation grant PHY-0244811 and the Department of
Energy garnt No. DE-FG01-00ER45832.


\begin{thebibliography}{99}

\bibitem{twodim} For reviews and references to original literature, see
I. Klebanov, {\tt hep-th9108019}, S.R. Das, {\tt hep-th/9211085}, 
A. Jevicki, {\tt hep-th/9309115}, P. Ginsparg and G. Moore,
{\tt hep-th/9304011}, J. Polchinski, {\tt hep-th/9411028.}

\bibitem{noncritical} S.R. Das, S. Naik and S.R. Wadia,
{\em Mod. Phys. Lett.} {\bf A4} (1989) 1033; 
J. Polchinski,
Nucl.Phys. {\em B324} (1989) 123;
S.R. Das, A. Dhar and
S.R. Wadia, {\em Mod. Phys. Lett.} {\bf A5} (1990) 799.

\bibitem{pola} J. Polchinski, 
{\em Nucl.Phys.} {\bf B346} (1990) 253.

\bibitem{dasjev} S.R.Das and A. Jevicki, {\em Mod. Phys. Lett.} {\bf A5}
(1990) 1639.

\bibitem{collective} A. Jevicki and B. Sakita,{\em Nucl. Phys.} {\bf B165}
(1980) 511.

\bibitem{fermions} D. Gross and I. Klebanov, {\em Nucl.Phys.} {\bf
B352} (1991) 671 ; A.M. Sengupta and S.R. Wadia, {\em
Int. J. Mod. Phys.} {\bf A6} (1991) 1961.

\bibitem{bhole} G. Mandal, A. Sengupta and S.R. Wadia, {\em Mod.Phys.Lett.}
{\bf A6} (1991) 1685; E. Witten, {\em Phys.Rev.} {\bf D44} (1991) 314.

\bibitem{twosides} G.Moore, {\em Nucl. Phys.} {\bf B368} (1992) 557;
G. Moore, R. Plesser and S. Ramgoolam, {\em Nucl. Phys. }{\bf B 377}
(1992) 143, {\tt hep-th/9111035}; J. Polchinski, {\tt hep-th/9411028};
A. Dhar, G. Mandal and S.R. Wadia, {\em Nucl. Phys.} {\bf B454} (1995)
{\tt hep-th/9507041}

\bibitem{liouvillebrane} V. Fateev, A.B. Zamolodchikov and
Al.B. Zamolodchikov, {\tt hep-th/0001012}; J. Teschner, {\tt
hep-th/0009138}; A.B. Zamolodchikov and Al. B. Zamolodchikov, {\tt
hep-th/0101152}.

\bibitem{verlinone} J. McGreevy and H. Verlinde, {\tt hep-th/0304224}

\bibitem{kms} I.R. Klebanov, J. Maldacena and N. Seiberg, {\em JHEP}
{\bf 0307} (2003) 045, {\tt hep-th/0305159}.

\bibitem{verlintwo} J. McGreevy, J. Teschner and H. Verlinde,
{\tt hep-th/0305194.}

\bibitem{zerob} T. Takayanagi and N. Toumbas, {\tt hep-th/0307083};
M. Douglas, I. Klebanov, D. Kutasov, J. Maldacena, E. Martinec and 
N. Seiberg, {\tt hep-th/0307195}

\bibitem{zeroa}  I.R. Klebanov, J. Maldacena and N. Seiberg,
{\tt hep-th/0309168}.

\bibitem{shenker} S. Shenker, in {\em Cargese 1990, Proceedings, 
Random surfaces and quantum gravity* 191-200.}

\bibitem{jevickiexact} A. Jevicki, {\em Nucl. Phys.} {\bf 376} (1992)
  75. 

\bibitem{mende} J. Lee and P. Mende,
{\em Phys.Lett.} {\bf B312} (1993) 433,
{\tt hep-th/9211049}.

\bibitem{dmwclass} A. Dhar, G. Mandal and S.R. Wadia, 
{\em Int.J.Mod.Phys.} {\bf A8} (1993) 3811,
{\tt hep-th/9212027.}

\bibitem{mwclass}. G. Mandal and S.R. Wadia, {\tt hep-th/0312192.}

\bibitem{sen} For a detailed discussion of this aspect see A. Sen,
{\tt hep-th/0308068}.

\bibitem{kraus} M. Gutperle and P. Kraus, {\tt hep-th/0308047.}

\bibitem{polchinski} J. Polchinski, {\em Nucl.Phys.} {\bf B362} (1991) 
125.

\bibitem{dmwbose} A. Dhar, G. Mandal and S.R. Wadia,
{\em Int.J.Mod.Phys.} {\bf A8} (1993) 325, {\tt hep-th/9204028};
{\em Mod.Phys.Lett.} {\bf A7} (1992) 3129, {\tt hep-th/9207011}.

\bibitem{dmathur} S.R. Das and S.D. Mathur, {\em Phys. Lett.} {\bf B
  365} (1996) 79, {\tt hep-th/9507141}

\bibitem{avanjevicki} J. Avan and A. Jevicki, {\em Nucl. Phys.}
{\bf B397} (1993) 672, {\tt hep-th/9209036.}; A. Jevicki,
{\tt hep-th/9302106}.

\bibitem{bosoni} For reviews see e.g. J. von Delft and H. Schoeller,
{\tt cond-mat/9805275}; S. Rao and D. Sen, {\tt cond-mat/0005492}.

\bibitem{deboer} J. de Boer, A. Sinkovics, E. Verlinde and J.T. Yee,
{\tt hep-th/0312135}

\bibitem{progress} S.R. Das, S.D. Mathur and P. Mukhopadhyay, {\em
  work in progress}


\end{thebibliography}
\end{document}